\documentclass[aps,prd,preprint,superscriptaddress,nofootinbib]{revtex4}
\usepackage{verbatim}
\usepackage{slashed}
\usepackage{epsfig}
\usepackage{graphicx}
\usepackage{subfigure}
\usepackage{amsmath}
\usepackage{mathrsfs}
\usepackage{bm}

\def\lQ{\Lambda_{\rm QCD}}
\def\als{\alpha_{\rm s}}

\begin{document}

\preprint{UB-ECM-10/39 $\;$ ICCUB-10-116}

\title{The discrete contribution to $\psi^{\prime}\to J/\psi+\gamma\gamma$}


\author{Zhi-Guo He}

\email{hzgzlh@ecm.ub.es}

\affiliation{\small{\it{Departament d'Estructura i Constituents de
la Mat\`eria
                  and Institut de Ci\`encies del Cosmos}}\\
        \small{\it{Universitat de Barcelona}}\\
        \small{\it{Diagonal, 647, E-08028 Barcelona, Catalonia, Spain.}}}

\author{Xiao-Rui Lu}

\email{xiaorui@gucas.ac.cn}

\address{Physics Department, Graduate University of Chinese Academy of Sciences,\\
Beijing, 100049, China}

\author{Joan Soto}
\email{joan.soto@ub.edu}
\affiliation{\small{\it{Departament d'Estructura i Constituents de
la Mat\`eria
                   and Institut de Ci\`encies del Cosmos}}\\
        \small{\it{Universitat de Barcelona}}\\
        \small{\it{Diagonal, 647, E-08028 Barcelona, Catalonia, Spain.}}}

\author{Yangheng Zheng}
\email{zhengyh@gucas.ac.cn}

\address{Physics Department, Graduate University of Chinese Academy of Sciences,\\
Beijing, 100049, China}



\date{\today}

\begin{abstract}
The decay mode $\psi(2S)\to J/\psi+\gamma\gamma$ is proposed
in order to experimentally identify the effects of the coupling of charmonium states
to the continuum $D\bar D$ states. To have a better understanding of such a two-photon
decay process, in this work we restrict ourselves to investigate the contribution of
the discrete part, in which the photons are mainly produced via the intermediate states
$\chi_{cJ}(nP)$. Besides calculating the resonance contributions of $\chi_{cJ}(1P)\; (J=0,1,2)$,
we also take into account the contributions of the higher excited states $\chi_{cJ}(2P)$
and the interference effect among the $1P$ and $2P$ states. We find that the contribution
of the $2P$ states and the interference terms to the total decay width is very tiny. However,
for specific regions of the Dalitz plot, off the resonance peaks, we find that these
contributions are sizable and should also be accounted for. We also provide the photon
spectrum and study the polarization of $J/\psi$.
\end{abstract}

\maketitle

\section{Introduction}

Electromagnetic processes have always provided invaluable probes of
the strong interactions, the most prominent example being the deep
inelastic scattering experiments in the early 1970s which
eventually established QCD as the fundamental theory of the strong
interactions. Among the current challenges of QCD is the description
of the plethora of charmoniumlike states discovered during the past
years, the so-called ``XYZ" (for reviews see, e.g.,
Refs. \cite{Swanson:2006st,Zhu:2007wz,Nielsen:2009uh,Brambilla:2010cs}),
that lie above the $D\bar D$ threshold, and do not fit potential
model expectations. It is widely accepted that the effects of the
coupling of a core charmonium, namely, a mainly $c\bar c$ bound
state, to a $D\bar D$ meson pair, the so-called coupled-channel
effects, are an important ingredient to understand those states. We
shall call the latter continuum states and the former valence
states. We will argue below that two-photon transitions between
heavy quarkonium states may provide important experimental
information on the continuum-valence coupling.

Historically, it has been recognized for a long time that the
continuum states may shift the mass spectrum of a pure $c\bar{c}$
state considerably
\cite{Eichten:1978tg,Eichten:1979ms,Heikkila:1983wd}. Recently,
exploratory investigations on the mixing between discrete and
continuum states in charmonium have been carried out in lattice QCD
\cite{Bali:2009er}. Lattice QCD has also provided a detailed study
of the so-called ``string breaking" in the static approximation
\cite{Bali:2005fu}, which may be used to extract information on the
continuum-valence coupling \cite{clark}. If the heavy quark mass $m$
is much larger than the remaining scales in the system, then the
heavy quarks move slowly in their center-of-mass frame, say with a
typical velocity $v_{Q}\ll 1$, that is, with typical three momentum
$mv_{Q}$ and hence with typical binding energy $mv_{Q}^{2}$.
Nonrelativistic QCD (NRQCD)
\cite{Caswell:1985ui,Thacker:1990bm,Bodwin:1994jh} can be used to
factorize the contributions from energies larger or of the order of
$m$, and provides a good starting point. Recall that the heavy-light
meson pair threshold lies at a typical nonrelativistic energy $\sim
\lQ$, according to heavy quark effective theory (HQET) counting rules
\cite{Neubert:1993mb}. Then, if the binding energy of a heavy
quarkonium state lies much below open flavor threshold, namely,
$mv_{Q}^{2}\ll \lQ$ one can integrate out energies of order $\sim
\lQ$. This leads to potential NRQCD (pNRQCD) in the strong coupling
regime \cite{Brambilla:1999xf}. The dynamics of this effective
theory reduces to a heavy quark and a heavy antiquark interacting
through a
potential\cite{Brambilla:2000gk,Pineda:2000sz,Brambilla:2003mu} (see
\cite{Brambilla:2004jw} for a review). Hence, in order to understand
the properties of these states, there is no need to introduce
explicitly the continuum states if the potential is chosen
appropriately. However, for states close to or above the open flavor
threshold, the coupling of continuum to valence states needs to be
addressed, and so far it is not known how to proceed in a model
independent way \footnote{An hadronic effective theory has recently
been introduced to study particular states very close to threshold,
most notably the $X(3872)$, which includes coupled channel effects
\cite{Artoisenet:2010va}. See also \cite{Guo:2010ak} for an even more recent proposal.}. Hence, most of the analysis has been
done by using different models. Recently, some general features of
the coupled-channel effects have been obtained in the quark model
\cite{Barnes:2007xu} under the assumptions that valence-continuum
coupling is described by the $^3P_0$ model \cite{Le Yaouanc:1972ae}
and the interaction between two mesons is negligible. These results
suggest that for the low-lying states the effect of continuum
channels is hidden in the parameters in the potential model. In Ref.
\cite{Barnes:2010gs}, it is mentioned that the radiative transition
process may be sensitive to the continuum components. However,
previous works,  based on Cornell coupled-channel formalism,
indicate that the relativistic corrections
\cite{Sebastian:1979gq,Li:2009zu} may be more important than the
continuum contributions \cite{Eichten:2004uh}. Therefore, it is not
straightforward to disentangle the continuum contributions in the
one-photon transition process.

Here, we propose a new process, namely, $\psi(2S)\to
J/\psi+\gamma\gamma$ which can provide an additional opportunity to
pin down the valence-continuum coupling. From the theoretical point
of view, electromagnetic transitions allow a cleaner analysis than
hadronic decay processes. At the amplitude level, the contribution
to this two-photon transition can be divided into two parts. We
refer to the first one as the discrete part, which involves
charmonium states only. The discrete part is dominated by the
following process: the $\psi(2S)$ state decays into a real or virtual
$\chi_{cJ}(nP)$ state by radiating one photon, and then, the real or
virtual $\chi_{cJ}(nP)$ state decays into $J/\psi$ plus another
photon. This process not only includes the cascade decay process
(on-shell region), but
also the off-shell region. We will study the whole phase space.
The second part is referred to as the continuum part, in which at
least one photon is emitted by an intermediate charmed meson. In the
discrete part, the $\chi_{cJ}(1P)(J=0,1,2)$ states can be on-shell,
so the total contribution of the discrete subprocess should be much
larger than that of the continuum part. However, when the invariant
mass of $J/\psi$ and one of the photons is far away from the
resonance regions of the $\chi_{cJ}(1P)(J=0,1,2)$ states, the
discrete part contribution will drop down very fast. Thus, the
contribution of the continuum part may be important and measurable
in the off-shell region. Let us mention that in Ref.
\cite{Guetta:1999vb}, a similar decay process of $B(D)^{\ast}\to
B(D)+2\gamma$ is suggested to determine the values of the strong
couplings $g_{_{B(D)^{\ast}B(D)\pi}}$ and
$g_{_{B(D)^{\ast}B(D)\gamma}}$.

On the experimental side, such a two-photon transition process has
already been studied in the 1970s and 1980s
\cite{Biddick:1977sv,Himel:1980fj}. In recent years, more precise
measurements were carried out by the BESIII
\cite{Bai:2004cg,Ablikim:2005yd} and CLEO \cite{Adam:2005uh,:2008kb}
Collaborations.
However, they focused on the investigation of the cascade decay
$\psi(2S)\to \chi_{cJ}+\gamma$ followed by $\chi_{cJ}\to
J/\psi+\gamma$ and on the study of the properties of the $\chi_{cJ}$
states. Only in Ref.\cite{:2008kb}, was a discussion made on the
possible amplitude of the two-photon transition in treating the
backgrounds of the $\chi_{cJ}$ states. Hence, so far, no one has
ever used it to study coupled-channel effects. Recently, the BESIII
\cite{Harris:2010zp} Collaboration reported the significant data
excess from the known cascade  backgrounds, which was interpreted
as the nonresonance decay of $\psi(2S)\to J/\psi\gamma\gamma$. We
remark that this measurement is very sensitive to the line shapes of
the $\chi_{cJ}$ states, especially in the data selection region.

Because of the above reason, the significance of an eventual
experimental determination of the continuous process depends very
much on our knowledge on the discrete states. As far as we know, the
discrete contribution has not been fully studied yet, and only the
individual contribution of each $\chi_{cJ}$ state is known by using
the non relativistic Breit-Wigner formula together with the
dynamical factors to describe the $\chi_{cJ}$ line shape in the
cascade decay of $\psi(2S)\to \chi_{cJ}+\gamma$ and $\chi_{cJ}\to
J/\psi+\gamma$ \cite{:2008kb}.
 So, in this paper
we restrict ourselves to analyze the discrete contribution to this
decay assuming that the coupling of $\psi'$ and $J/\psi$ to $D\bar
D$ is zero, and leave for a future work, the detailed evaluation
of $D\bar D$ meson pair loops effects.

 In this work, we will use effective field theory methods to calculate
the decay width, the photon spectrum, and the $J/\psi$ polarization in the discrete
subprocess. A complete study of the whole contribution of the
discrete $\chi_{cJ}(1P)$ states together with some higher radial
excitations, $\chi_{cJ}(2P)$\footnote{The contribution of higher
$nP$ states, where $n\geq 3$, are ignored.}, will be carried out.
The rest of this paper is organized as followings: in Sec.II, we
will briefly introduce the effective Lagrangian we use and determine
the value of the effective couplings; in Sec.III, we will calculate
the discrete contribution to the $\psi(2S)\to J/\psi+\gamma\gamma$
process and show the results, and the discussion and summary will be
given in the last section.

\section{Effective Lagrangian For Radiative Transitions}

The heavy quarkonium states are mainly constituted by a $Q\bar{Q}$
pair and classified according to the spectroscopic notation
$n^{2S+1}L_{J}$ , where $n=1,2,\ldots$ is the radial quantum number,
$S=0,1$ is the total spin of the heavy quark pair, $L=0,1,2\ldots$
(or $S,P,D\ldots$) is the orbital angular momentum, and $J$ is the
total angular momentum. They have parity $P=(-1)^{L+1}$ and charge
conjugation $C=(-1)^{L+S}$.

As mentioned in the introduction, NRQCD is a good starting point to
describe this system. The LO NRQCD Lagrangian is invariant under
S=$SU(2)_Q\otimes SU(2)_{\bar Q}$ spin-symmetry group, an
approximate symmetry of the heavy quarkonium states, that is
inherited in the subsequent effective theories. We assume that the
entire dynamics of these states can be described by a
(nonperturbative) potential. That is the case if they are in the
strong coupling regime of pNRQCD. This is a reasonable assumption
for the $\chi_{cJ}(1P)$ states \cite{Brambilla:2001xy}, and to
lesser extend for $J/\psi$\footnote{There are indications that
$J/\psi$ may well be better described by the weak coupling regime of
pNRQCD \cite{Penin:2004ay,Brambilla:2005zw,Kiyo:2010zz}. At leading
order in this regime, the dynamics is described by a (perturbative)
potential, and the multipole expansion also holds.}. The remaining
states are close to or above threshold and are subject to the
uncertainties due to the influence of $D\bar D$ pairs, that we plan
to analyze in a separate work. In any case, we will only use the
fact that the typical energy of the emitted photons is at the
ultrasoft scale $mv_{Q}^{2}$, and the typical size of the system at
the soft scale $\frac{1}{mv_{Q}}$. This allows hs to carry out the
multipole expansion of the photon field about the center-of-mass
coordinate, which means that the photons see the heavy quarkonium
states as pointlike particles. Hence, it is most convenient to
introduce hadronic spin-symmetry multiplets, in an analogous way as
it was initially done in HQET \cite{Neubert:1993mb}.

For heavy quarkonium states, this formalism was developed in
Ref.\cite{Casalbuoni:1992yd}. The states have the same radial number
$n$ and the same orbital momentum $L$ can also be expressed by means
of a single multiplet
$J^{\mu_1\ldots\mu_L}$
\cite{Casalbuoni:1992yd},
\begin{eqnarray}
J^{\mu_1\ldots\mu_L}&=&\frac{1+\slashed{v}}{2}(H_{L+1}^{\mu_1\ldots\mu_L\alpha}\gamma_{\alpha}+\frac{1}{\sqrt{L(L+1)}}
\sum_{i=1}^{L}\epsilon^{\mu_i\alpha\beta\gamma}v_{\alpha}\gamma_{\beta}H_{L\gamma}^{\mu_1\ldots\mu_{i-1}\mu_{i+1}\ldots\mu_L}
\nonumber\\&+&
\frac{1}{L}\sqrt{\frac{2L-1}{2L+1}}\sum_{i=1}^{L}(\gamma^{\mu_i}-v^{\mu_i})H_{L-1}^{\mu_1\ldots\mu_{i-1}\mu_{i+1}\ldots\mu_L}
\nonumber\\&-&
\frac{2}{L\sqrt{(2L-1)(2L+1)}}\sum_{i<j}(g^{\mu_i\mu_j}-v^{\mu_i}v^{\mu_j})\gamma_{\alpha}
H_{L-1}^{\alpha\mu_1\ldots\mu_{i-1}\mu_{i+1}\ldots\mu_{j-1}\mu_{j+1}\ldots\mu_L}
\nonumber\\&+&K_{L}^{\mu_1\ldots\mu_{L}}\gamma^{5})\frac{1-\slashed{v}}{2}
\end{eqnarray}
where $v^{\mu}$ is the four velocity associated to the multiplet
$J^{\mu_1\ldots\mu_L}$ (not to be mistaken by $v_{Q}$, the typical
velocity of the heavy quark in the heavy quarkonium rest frame),
$K_{L}^{\mu_1\ldots\mu_{L}}$ represents the spin-singlet effective
field, and
$H_{L-1}^{\mu_1\ldots\mu_{L-1}}$,$H_{L}^{\mu_1\ldots\mu_{L}}$ and
$H_{L+1}^{\mu_1\ldots\mu_{L+1}}$ represent the three spin-triplet
effective fields with $J=L-1,L$, and $L+1$, respectively. The four
tensors are all completely symmetric and traceless and satisfy the
transverse condition
\begin{equation}\label{trans}
v_{\mu_i} K_{L}^{\mu_1\ldots\mu_i\ldots\mu_{L}}=0\quad ,\quad  v_{\mu_j} H_{J}^{\mu_1\ldots\mu_j\ldots\mu_{J}}=0,
\end{equation}
$i=1,\dots , L$, $j=1,\dots , J$. The properties of $H$ and $K$ under parity,
charge conjugation and heavy quark spin transformations can be easily obtained
by assuming that the corresponding transformation rules of the multiplet
$J^{\mu_1\ldots\mu_L}$ follow as:
\begin{subequations}
\begin{eqnarray}
J^{\mu_1\ldots\mu_L}\stackrel{P}{\longrightarrow}\gamma^{0}J_{\mu_1\ldots\mu_L}\gamma^{0},
v^{\mu}\stackrel{P}{\longrightarrow}v_{\mu},
\end{eqnarray}
\begin{eqnarray}
J^{\mu_1\ldots\mu_L}\stackrel{C}{\longrightarrow}(-1)^{L+1}C[J_{\mu_1\ldots\mu_L}]^{T}C,
\end{eqnarray}
\begin{eqnarray}
J^{\mu_1\ldots\mu_L}\stackrel{\rm S}{\longrightarrow}SJ_{\mu_1\ldots\mu_L}S^{\prime\dagger},
\end{eqnarray}
\end{subequations}
where $C$ is the charge conjugation matrix ($C=i\gamma^{2}\gamma^{0}$ in the Dirac representation), 
and $S\in SU(2)_Q$ and $S^{\prime} \in SU(2)_{\bar Q}$ correspond to the heavy quark and
heavy antiquark spin-symmetry groups ($[S,\slashed{v}]=[S^{\prime},\slashed{v}]=0$).

Since we are going to consider the two-photon decay of $\psi(2S)$
into $\psi(1S)$ via the intermediate states $\chi_{cJ}(nP)$, it
will be helpful to give the explicit expressions of the $S$- and
$P$-wave multiplets that follow from Eq.(1). For the $L=S$ case, we have
\begin{equation}
J=\frac{1+\slashed{v}}{2}(H_1^{\mu}\gamma_{\mu}-K_0\gamma^{5})\frac{1-\slashed{v}}{2},
\end{equation}
and for the $L=P$ case,
\begin{equation}
J^{\mu }=\frac{1+\slashed{v}}{2}\Big\{
H_2^{\mu \alpha } \gamma_\alpha  + {1 \over \sqrt{2}}
\epsilon^{\mu \alpha \beta \gamma} v_\alpha \gamma_\beta H_{1
\gamma}+\frac{1}{\sqrt{3}}
(\gamma^{\mu} -v^{\mu}) H_0  + K_1^{\mu }\gamma_5 \Big\}\frac{1-\slashed{v}}{2}.
\end{equation}
The radiative transitions between $mS$ and $nP$ charmonium states in the
nonrelativistic limit is given by the Lagrangian:
\begin{equation}
\mathcal{L}=\sum_{m,n}\delta^{nP,mS}\mathrm{Tr}[\bar{J}({mS})J_{\mu}(nP)]v_{\nu}F^{\mu\nu}+\mathrm{H.c.},
\label{vertex}
\end{equation}
where $\delta^{nP,mS}$ is the coupling constant, and $F^{\mu\nu}$ is the
electromagnetic tensor. The Lagrangian in Eq.(6) preserves parity, charge
conjugation, gauge invariance, and heavy quark and antiquark spin symmetry.

Using the effective Lagrangian, it is straightforward
to calculate the E1 transition decay widths:
\begin{subequations}\label{E1}
\begin{equation}
\Gamma(m^3S_1\to n ^3P_J)=(2J+1)\frac{(\delta_{J}^{nP,mS})^2}{144\pi}k_{\gamma}^{3}
\frac{(M_{mS}+M_{nP})^4}{M_{mS}^3M_{nP}};
\end{equation}
\begin{equation}
\Gamma(n^3P_J\to m ^3S_1)=\frac{(\delta_{J}^{nP,mS})^2}{48\pi}k_{\gamma}^{3}
\frac{(M_{mS}+M_{nP})^4}{M_{mS}M_{nP}^3};
\end{equation}
\label{decays}
\end{subequations}
where $k_{\gamma}$ is the energy of the emitted photon. The results
are slightly different from those given in
Refs.\cite{Casalbuoni:1992yd,DeFazio:2008xq}. It is because the
initial and final charmonium states can not be static
simultaneously, and we choose different values of $v_\mu$ for them
to maintain the transverse condition in Eq.(\ref{trans}). Namely, in
the vertex (\ref{vertex}) we have substituted the velocity $v^\mu$
in
the current that produces the outgoing particle by $v^{\mu}_{f}$,
its four velocity. The remaining velocities (i.e. the explicit one
and the one in the current that annihilates the incoming particle)
are chosen as $v^{\mu}_i$, the four velocity of the
incoming particle. The explicit calculations in (\ref{decays}) show that
these changes maintain the spin-symmetry ratios between the decay widths
of the different states ( provided that the same mass is used for each
spin multiplet ).
Note that $v_{i}.v_{f}=1+{\cal O} (v_{Q}^4)$, and
the results of
Ref.\cite{
DeFazio:2008xq} differ from ours by
${\cal O} (v_{Q}^4)$ contributions. In (\ref{decays}) we have also
allowed the coupling constant $\delta^{nP,mS}$ of the spin-symmetry
multiplet to depend on the total angular momentum $J$, namely,
$\delta^{nP,mS} \to \delta^{nP,mS}_J = \delta^{nP,mS} + {\cal O}
(1/m^2)$. In this way we are accounting for spin-symmetry breaking
terms due to the spin-orbit, spin-spin and tensor potentials in pNRQCD
(or in potential models). Note that although spin-breaking terms look
like an ${\cal O}(1/m)$ in NRQCD (like in HQET), this is actually the
case only when ultrasoft particles are emitted, for instance, in magnetic
transitions, or in processes involving pseudo-Goldstone bosons \cite{Guo:2010ak}.
Soft emissions can only be virtual and lead to the above mentioned ${\cal O}(1/m^2)$
spin-breaking potentials, which make the coupling constants different at this order.
In conventional potential models, the differences
between $\delta^{nP,mS}_J$ are usually taken to be of order $v_{Q}^2$, although
on general grounds they could be as large as order $v_{Q}$
\cite{Pineda:2000sz}. $\delta^{nP,mS}$ can, in principle, be obtained
by calculating the matrix element of the electromagnetic current
between the wave functions of the $nP$ and $mS$ states in pNRQCD (or
in any potential model, see Ref.\cite{Eichten:2007qx} for a recent
review). Nowadays they can also be obtained from lattice
QCD \cite{Dudek:2006ej,Dudek:2009kk}. Here we will determine their
values using experimental data, except for those related to the
$\chi_{cJ}(2P)$ for which experimental data is not available. In the
last case the values will be estimated using a potential model.

For the $1P$ case, the $\psi(2S)\to \chi_{cJ}(1P)+\gamma$ and
$\chi_{cJ}(1P)\to J/\psi+\gamma$ processes have been measured with a
very high precision \cite{Nakamura:2010px}, so we can determine the
values of the relevant coupling constants by comparing with the
experimental results. Using the PDG data
\cite{Nakamura:2010px}, $M_{2S}=3.686$GeV, $M_{\psi(1S)}=3.097$GeV,
$M_{\chi_{cJ}}=3.415, 3.511,3.556 \;
\mathrm{GeV}\;(\mathrm{for}\; J=0,1,2)$,
$\Gamma_{\psi(2S)}=304\mathrm{keV}$,
$\Gamma_{\chi_{cJ}}=10.3,0.86,1.97\;\mathrm{MeV}\; (\mathrm{for}\; J=0,1,2),$
and
\begin{subequations}
\begin{equation}
\mathrm{Br}(\psi(2S)\to \chi_{c0}+\gamma)=9.62\%\,;\quad\mathrm{Br}(\chi_{c0}\to J/\psi+\gamma)=1.16\%;
\end{equation}
\begin{equation}
\mathrm{Br}(\psi(2S)\to \chi_{c1}+\gamma)=9.2\%\,;\quad\mathrm{Br}(\chi_{c1}\to J/\psi+\gamma)=34.4\%;
\end{equation}
\begin{equation}
\mathrm{Br}(\psi(2S)\to \chi_{c2}+\gamma)=8.74\%\,;\quad\mathrm{Br}(\chi_{c2}\to J/\psi+\gamma)=19.5\%;
\end{equation}
\end{subequations}
we obtain the absolute values of $\delta_{J}^{1P,2S}$ and
$\delta_{J}^{1P,1S}$ listed in Table I. For the $2P$ case, only the
$\chi_{c2}(2P)$ [formerly called $Z(3930)$\cite{Uehara:2005qd},] has been
included in the PDG, so we estimate the relevant parameters with the
help of a potential model. The spectrum of the $2P$ states and their
radiative decay into the lower $S$-wave states have been calculated by
numerous groups (for reviews see, e.g., Ref.\cite{Eichten:2007qx}). By
using the screened potential, the updated results of the charmonium
spectrum and the E1 transition rates are given in
Ref.\cite{Li:2009zu}. In this work, besides setting
$M_{\chi_{c2}(2P)}=3.929\mathrm{GeV}$,  $X(3872)$ is assigned to the
$\chi_{c1}(2P)$ state and $M_{\chi_{c0}(2P)}=3.842\mathrm{GeV}$ is
predicted. From the nonrelativistic results of
$\Gamma(\chi_{cJ}(2P)\to J/\psi(\psi(2S))+\gamma)$ presented in
Table IV of Ref.\cite{Li:2009zu}, we obtain the absolute values of
the corresponding coupling constants and give them in Table I.
Because the mass differences between the $2P$ states are much smaller
than the mass gap between the $2P$ and $2S$ ($1S$) states, in practice we
use the center of gravity mass
\begin{equation}
M_{(2P)}=\frac{M_{\chi_{c0}(2P)}+3\times M_{\chi_{c1}(2P)}+5\times M_{\chi_{c2}(2P)}}{9}
\end{equation}
and the center of gravity coupling constant
\begin{equation}
\delta^{2P,ms}\equiv\frac{\delta_0^{2P,mS}+3\times\delta_1^{2P,mS}+5\times\delta_2^{2P,mS}}{9}
\end{equation}
rather than calculating the contribution of the individual $2P$ states.

\begin{center}
\begin{table}
\caption{The numerical values of the coupling constants $\delta_{J}^{nP,mS} (\mathrm{GeV}^{-1})$
are shown. For the $n=1$ case, the results are obtained by fitting the experimental data, and for
$n=2$, the results are determined by comparing with the potential model predictions \cite{Li:2009zu}.}
\vspace{.3cm}
\begin{tabular}{|c|c|c|c|c|c|c|}
     \hline
     & $\chi_{c0}(1P)$&$\chi_{c1}(1P)$&$\chi_{c2}(1P)$&$\chi_{c0}(2P)$&$\chi_{c1}(2P)$&$\chi_{c2}(2P)$\\
        \hline
     $J/\psi$&0.211&0.230&0.228&$5.27\times10^{-2}$&$5.30\times10^{-2}$&$5.34\times10^{-2}$\\
        \hline
     $\psi(2S)$&0.224&0.235&0.273&0.410&0.413&0.416\\
        \hline
\end{tabular}
\end{table}
\end{center}

\section{Discrete Contribution to $\psi(2S)\to J/\psi+2\gamma$}

\begin{figure}
\begin{center}
\includegraphics[scale=0.8]{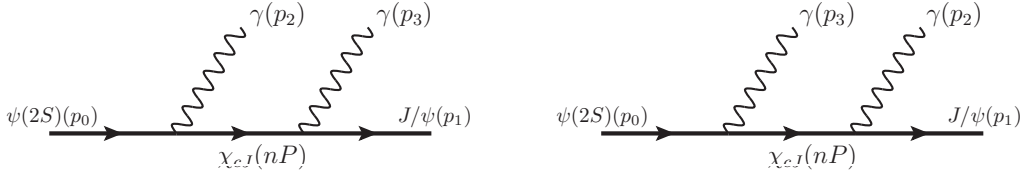}
\caption{The Feynman diagrams for $\psi(2S)$ decay into $J/\psi+2\gamma$ via intermediate states $\chi_{cJ}(nP)$.}
\label{fig:Feynman}
\end{center}
\end{figure}

Now, we proceed to calculate the decay rate of the process
$\psi(2S)(p_0)\to J/\psi(p_1)+\gamma(p_2)+\gamma(p_3)$ via intermediate states $\chi_{cJ}(nP)$.
Such $1\to 3$ process can be described by the following dimensionless variables:
\begin{equation}
x_i=\frac{2p_{0}\cdot p_i}{M_{\psi(2S)}^2},\;\sum_{i} x_{i}=2,
\end{equation}
In terms of $x_i$, the three-body phase space $\Phi_{(3)}$ can be written as
\begin{eqnarray}
\mathrm{d}\Phi_{(3)}=\frac{M_{\psi(2S)}^2}{2(4\pi)^3}
\delta(2-x_1-x_2-x_3) d x_1 d x_2 d x_3,
\end{eqnarray}
For each intermediate $\chi_{cJ}(nP)$, there are two Feynman diagrams, which are shown
in Fig.1. The corresponding Feynman amplitude is denoted by $\mathcal{M}^{\chi_{cJ}(nP)}$.
Putting the contributions of the three $2P$ states together, the total Feynman amplitude
is then divided into four parts:
\begin{equation}
\mathcal{M}^{\mathrm{Tot}}=\mathcal{M}^{\chi_{c0}(1P)}+\mathcal{M}^{\chi_{c1}(1P)}+\mathcal{M}^{\chi_{c2}(1P)}
+\mathcal{M}^{\chi_c(2P)}.
\end{equation}
Each of $\mathcal{M}$ on the right-hand side of the above equation can be obtained
from the Lagrangian in Eq.(6) upon making the same replacements as discussed after
(\ref{vertex}). Since the propagator of the $1P$ fields may become on-shell, self-energy
corrections must be considered. We approximate them by introducing a constant decay width,
which, parametrically, is ${\cal O}(m \als^2 v_{Q}^5 , \, m \alpha v_{Q}^4)$. Being
that these figures are much smaller than $mv_{Q}^2$, $1/m$ corrections to the static
propagator must be considered in order to match its size. As an alternative, we use
relativistic propagators, that include them. For example, the Feynman amplitude of the
first diagram in \figurename~\ref{fig:Feynman} for $\mathcal{M}^{\chi_{c2}(1P)}$, which
is the most complicated one, is
\begin{eqnarray}
\mathcal{M}^{\chi_{c2}(1P)}&=&\delta_2^{1P,1S}\delta_2^{1P,2S}
\mathrm{Tr}[\frac{1+\slashed{v}_{2S}}{2}\slashed{\epsilon}_{2S}\gamma^{\alpha}\frac{1+\slashed{v}_{p}}{2}]
\times\frac{(\Pi_{\alpha\alpha_1}\Pi_{\mu\mu_1}+\Pi_{\alpha\mu_1}\Pi_{\mu\alpha_1})/2-
\Pi_{\alpha\mu}\Pi_{\alpha_1\mu_1}/3}{v_p^2-1+\mathrm{I}*\Gamma_{\chi_{c2}(1P)}/M_{\chi_{c2}(1P)}}
\nonumber\\ &\times&
\mathrm{Tr}[\frac{1+\slashed{v}_{p}}{2}\gamma^{\alpha_1}\slashed{\epsilon}^{\ast}_{1S}\frac{1+\slashed{v}_{1S}}{2}]
\times
v_{\nu,2S}v_{\nu,p}\bar{F}_{2}^{\mu\nu}\bar{F}_{3}^{\mu_1\nu_1}.
\label{amplitude}
\end{eqnarray}
where $\Pi^{\alpha\beta}=(-g^{\alpha\beta}+v_{p}^{\alpha}v_{p}^{\beta})$,
$\bar{F}_{i}^{\alpha\beta}=p_{i}^{\alpha}\epsilon_{i}^{\ast\beta}-p_{i}^{\beta}\epsilon_{i}^{\ast\alpha}$,
$v_p^{\mu}=(p_0^{\mu}-p_2^{\mu})/M_{\chi_{c2}(1P)}$, and $\Gamma_{\chi_{c2}(1P)}$ is the total width of $\chi_{c2}(1P)$.
In the above expression, we have omitted the imaginary unit I, which is a global
factor and has no influence on the final result. For the convenience of further
discussion, we also divided the decay width of the discrete part into four parts:
\begin{eqnarray}
\Gamma_{\mathrm{dis}}(\psi(2S)\to J/\psi+\gamma\gamma)=
\Gamma^{1P}_{\rm{Ind}}+\Gamma^{1P}_{\mathrm{Int}}+\Gamma^{2P}\pm\Gamma^{1,2P}_{\mathrm{Int}},
\label{decay}
\end{eqnarray}
where $\Gamma^{1P}_{\rm{Ind}}$ is the sum of the three individual
contributions of the $\chi_{cJ}$ states, which is proportional to
$\sum_{J=0}^2 \vert\mathcal{M}^{\chi_{cJ}(1P)}\vert^2$,
$\Gamma^{1P}_{\mathrm{Int}}$ is the interference between the $1P$
states $(\sim \Re
\{\sum_{J\not=J'}\mathcal{M^\ast}^{\chi_{cJ}(1P)}\mathcal{M}^{\chi_{cJ'}(1P)}\})$,
$\Gamma^{2P}$ is the contribution involving the $2P$ states only,
and $\Gamma^{1,2P}_{\mathrm{Int}}$ is the interference contribution
between the $1P$ and $2P$ states.
 The ``$+(-)$" sign before the last term corresponds to the two possible
relative phase angles $0$ $(\pi)$ between the $1P$ and $2P$ states.

\begin{figure}
\begin{center}
\subfigure[]
{\includegraphics[width=0.48\textwidth]{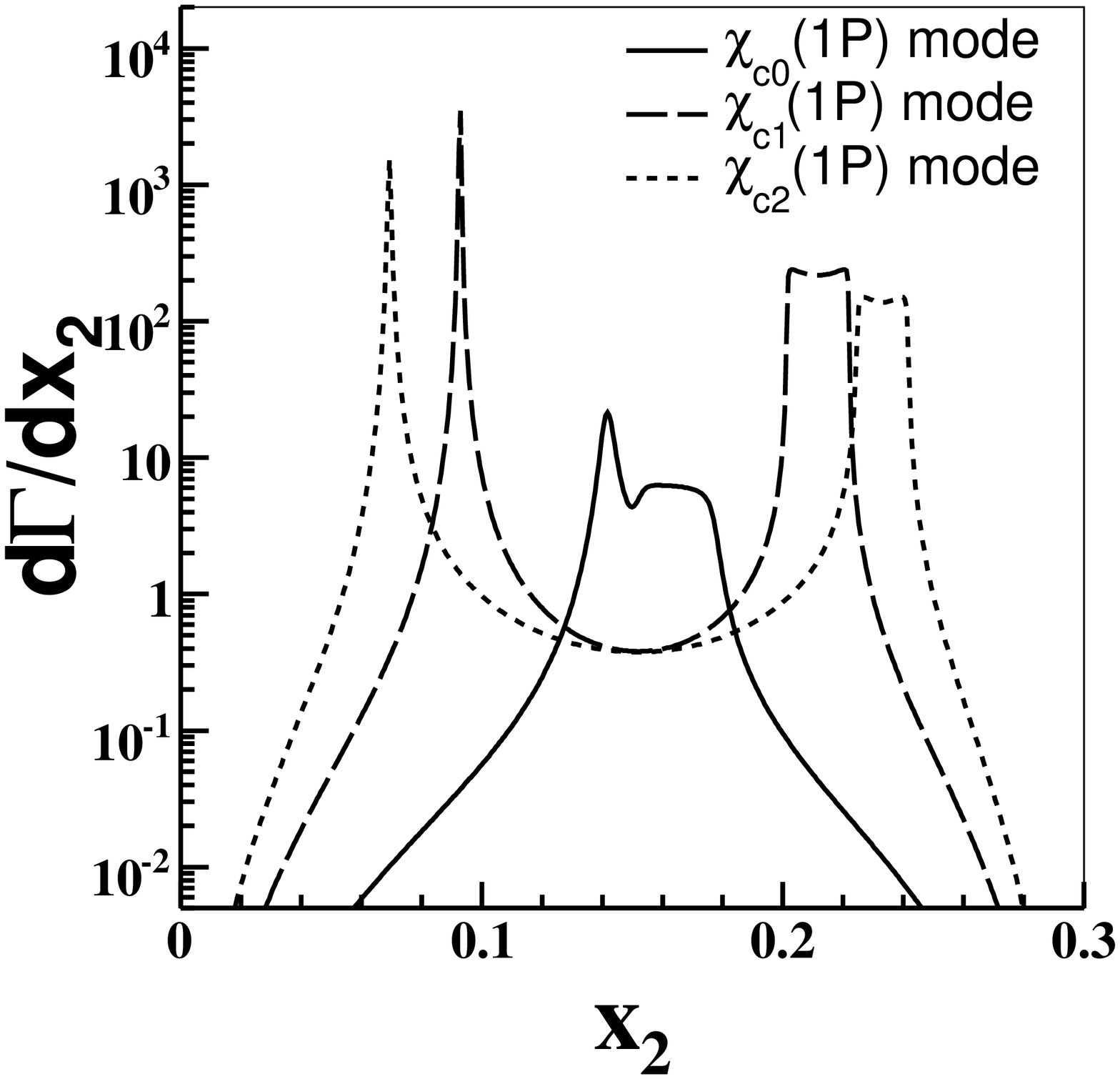}\label{fig:res_a}}
\subfigure[]
{\includegraphics[width=0.48\textwidth]{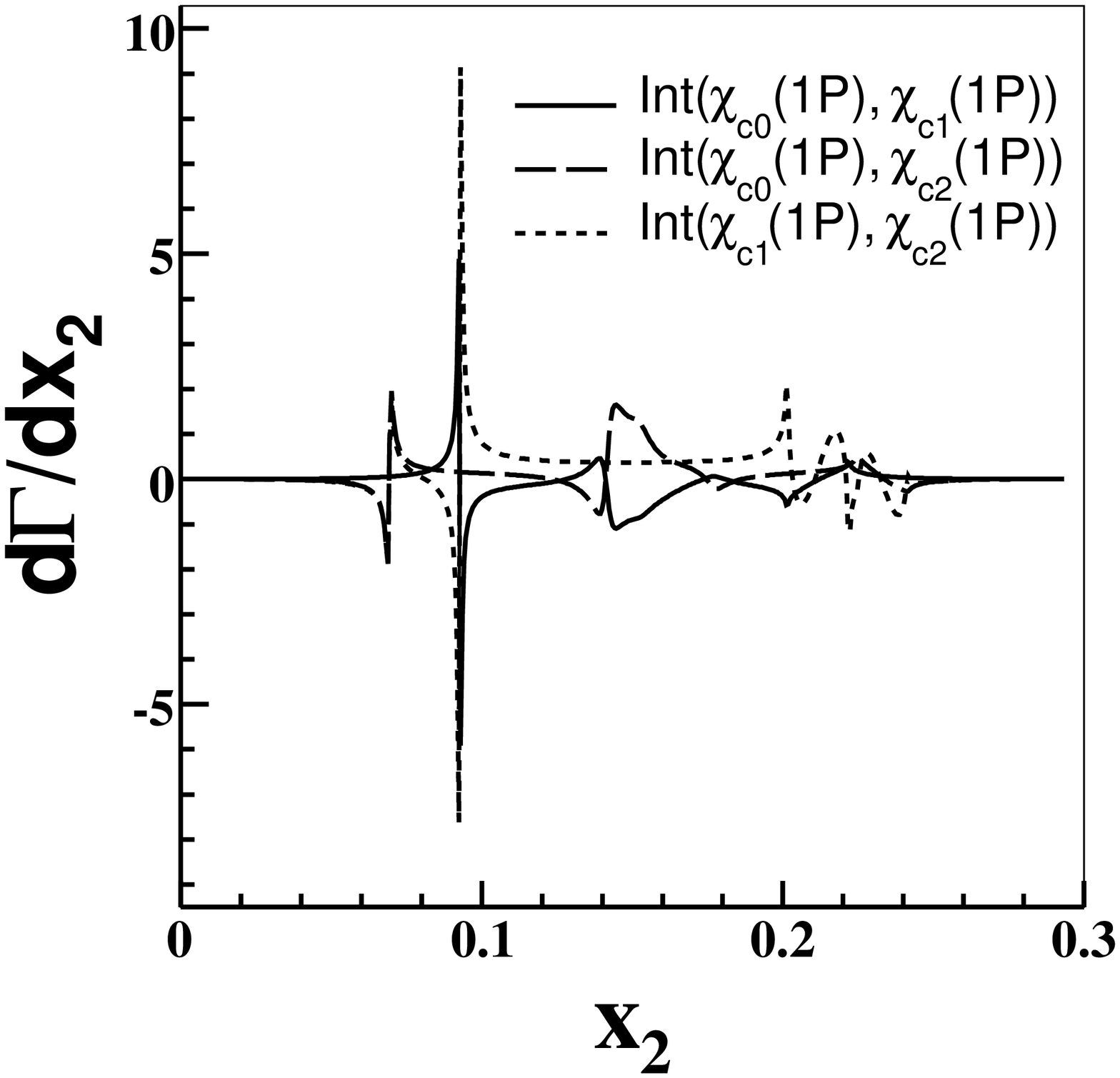}}
\caption{The partial decay width as a function of the photon energy
fraction $x_2$: (a) the individual contribution of the three
$\chi_{cJ}$ ($J$=0,1,2) states, corresponding to $\Gamma^{1P}_{\rm{Ind}}$ in (\ref{decay}), (b) the contribution of the interference terms
between the three $\chi_{cJ}$ ($J$=0,1,2) states, corresponding to $\Gamma^{1P}_{\rm{Int}}$ in (\ref{decay}).}
\label{fig:chic1P}
\end{center}
\end{figure}

\begin{figure}
\begin{center}
\subfigure[]
{\includegraphics[width=0.48\textwidth]{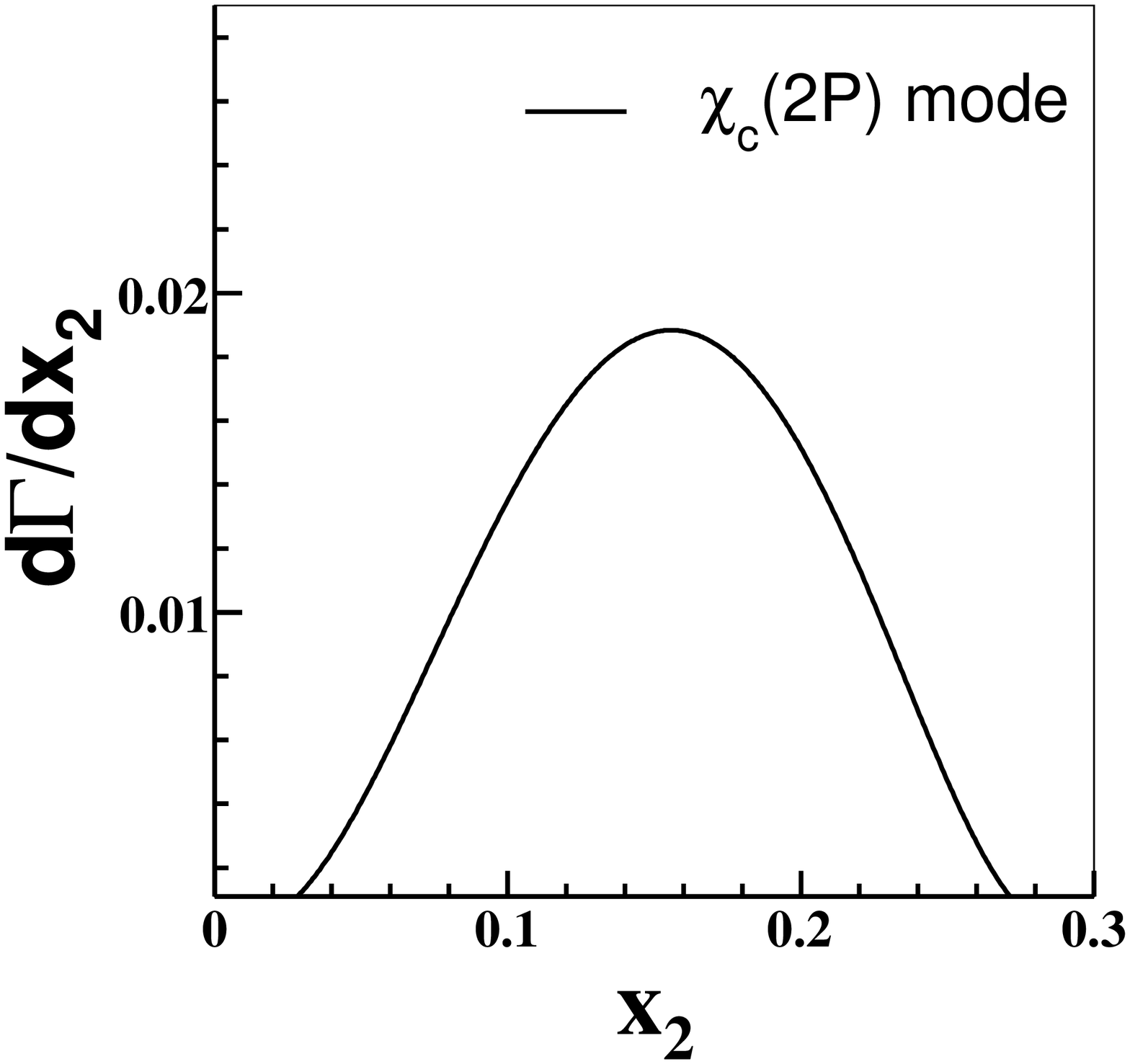}}
\subfigure[]
{\includegraphics[width=0.48\textwidth]{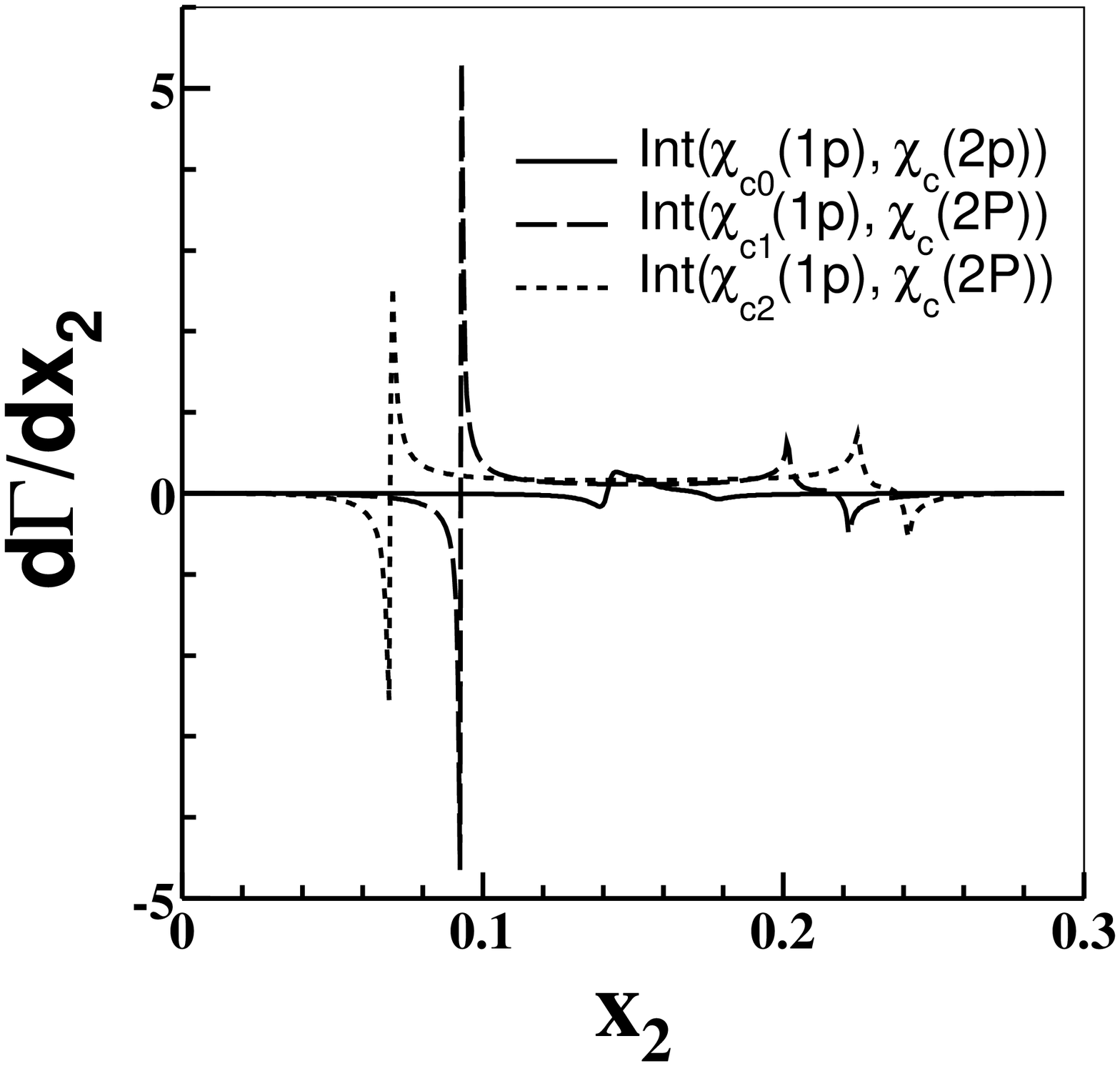}}
\caption{The partial decay width as a function of the photon energy
fraction $x_2$: (a) the contribution of the $2P$ states, corresponding to $\Gamma^{2P}$ in (\ref{decay}), (b)
 the contribution of the interference terms between the $2P$ and the three $1P$ states, corresponding to $\Gamma^{1,2P}_{\rm{Int}}$ in (\ref{decay}).}
\label{fig:chic2P}
\end{center}
\end{figure}

We have computed $\sum|\mathcal{M}^{Tot}|^2$ analytically, but the
outcome is too lengthy to be presented here. After doing the phase
space integrals numerically, we obtain that
\begin{eqnarray}
\Gamma^{1P}_{\rm{Ind}}=15.14\mathrm{keV}\simeq\sum_{J}\Gamma(\psi(2S)\to
\gamma+\chi_{cJ})\times\mathrm{Br}(\chi_{cJ}\to J/\psi+\gamma),
\nonumber \\
\Gamma^{1P}_{\mathrm{Int}}=5.95\times10^{-2}\mathrm{keV},\; \Gamma^{2P}=2.80\times10^{-3}\mathrm{keV},\;
\Gamma^{1,2P}_{\mathrm{Int}}=4.13\times10^{-2}\mathrm{keV}.
\label{total}
\end{eqnarray}
The numerical results yield that on the total decay width level the
effects of the interference among the $\chi_{cJ}(1P)$ states as well
as the effect of the $2P$ states are so small that they can be neglected.
We have also calculated the photon spectrum $\mathrm{d}\Phi_{(3)}/dx_2$,
and display the figures for each part separately in \figurename~\ref{fig:chic1P}
and  \figurename~\ref{fig:chic2P}, respectively.

Since we have chosen different $v$ for the initial and final states
in Eq.(\ref{amplitude}), we may also have a nontrivial $J/\psi$ polarization,
which should be zero in the single $v$ case. Similar
to the production case \cite{Gong:2008sn}, here, we define the polarization
parameter $\alpha$ as
\begin{eqnarray}
\alpha=\frac{\Gamma_{T}-2\Gamma_{L}}{\Gamma_{T}+2\Gamma_{L}},
\end{eqnarray}
where $\Gamma_{T}$ and $\Gamma_{L}$ are the decay widths for $J/\psi$ in
the transverse and longitudinal polarization states, respectively. We calculate
$\alpha$ in the rest frame of $\psi(2S)$ and get $\alpha\simeq-0.16$. This
is slightly different from zero, and we interpret it as a purely kinematic effect.
We find that $\alpha$ is mainly determined by the resonance contribution, being
the influence of the interference and the $2P$ terms tiny.

\begin{figure}
\begin{center}
\subfigure[]
{\includegraphics[width=0.32\textwidth]{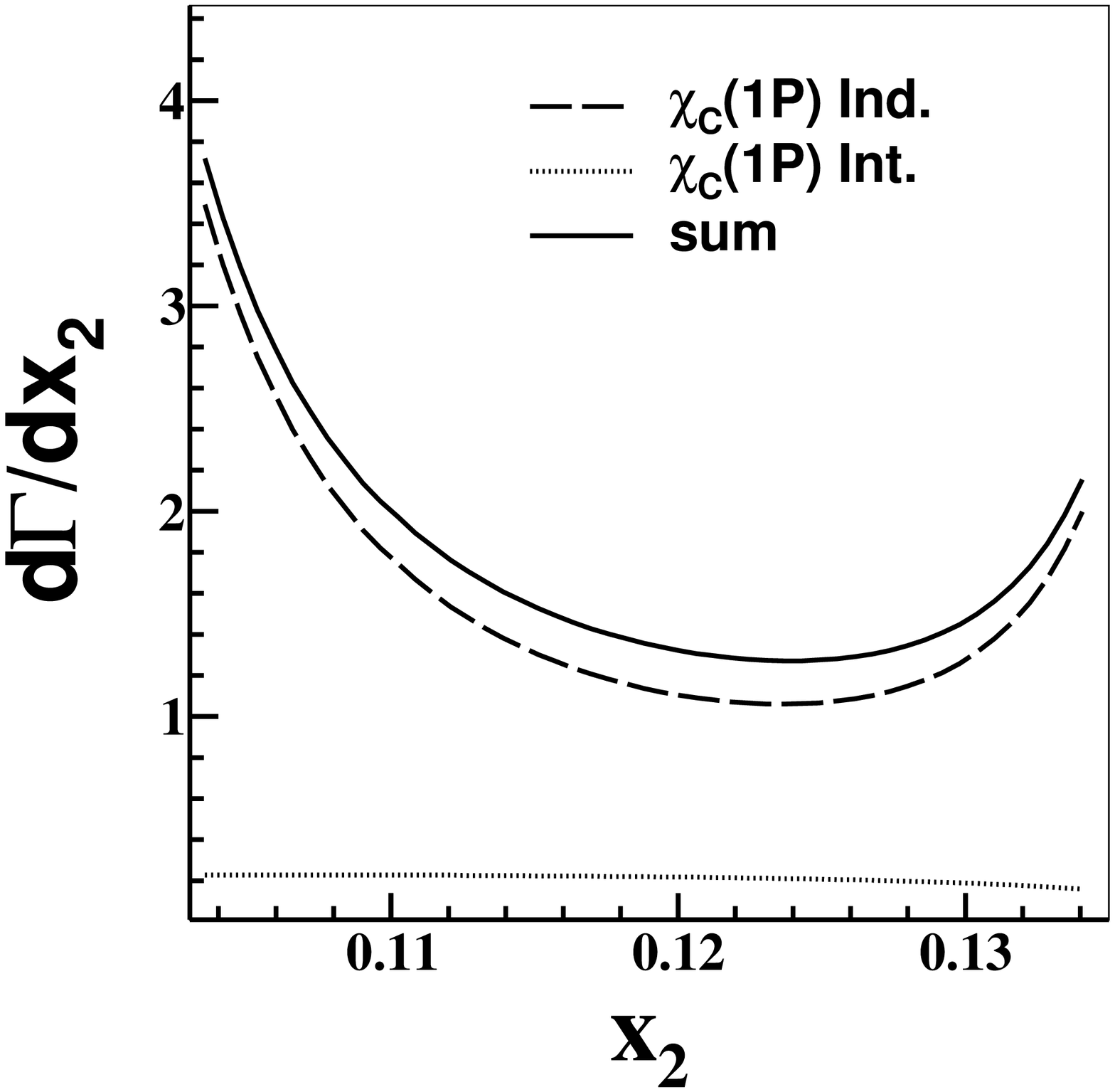}}
\subfigure[]
{\includegraphics[width=0.32\textwidth]{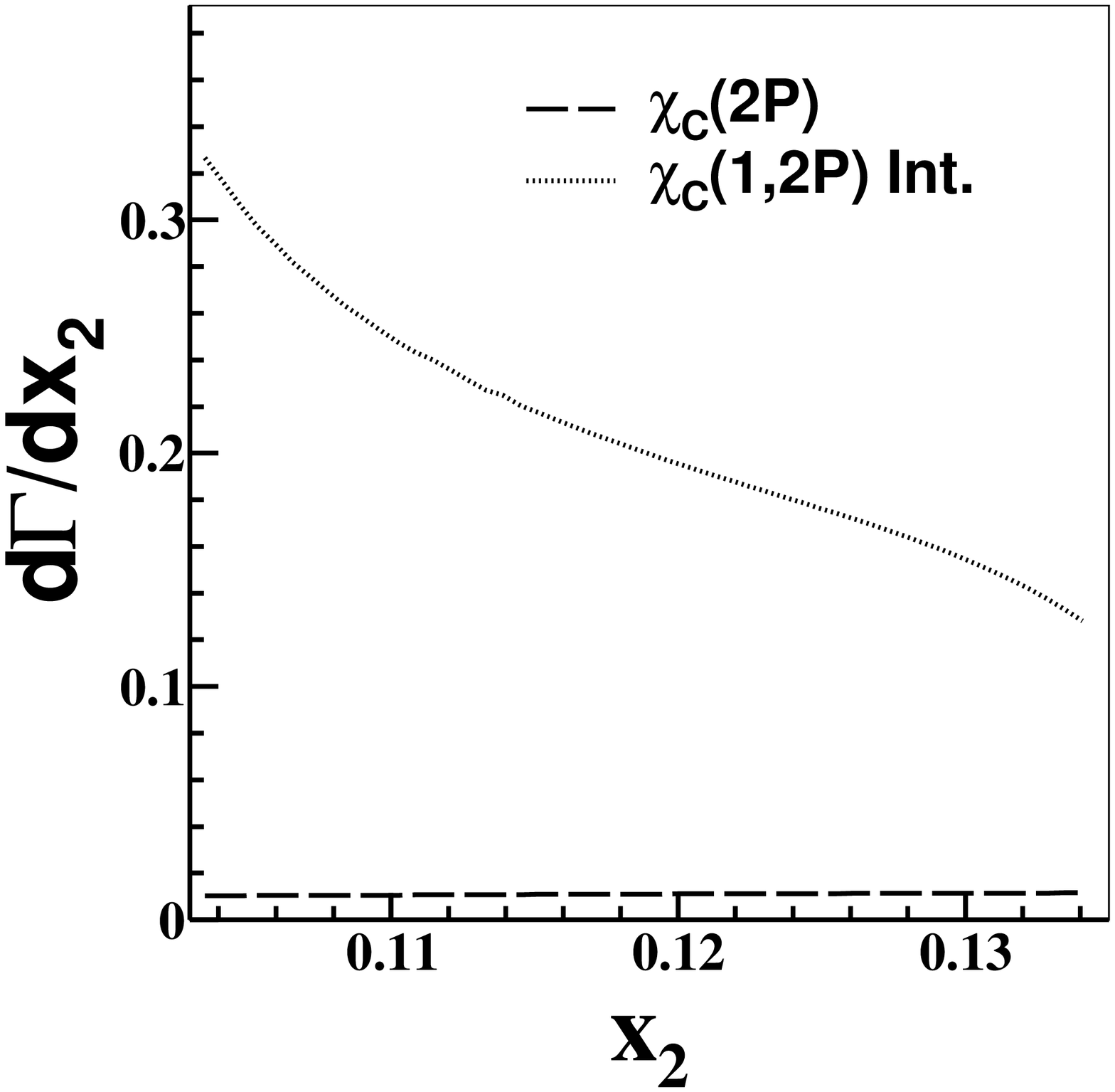}}
\subfigure[]
{\includegraphics[width=0.32\textwidth]{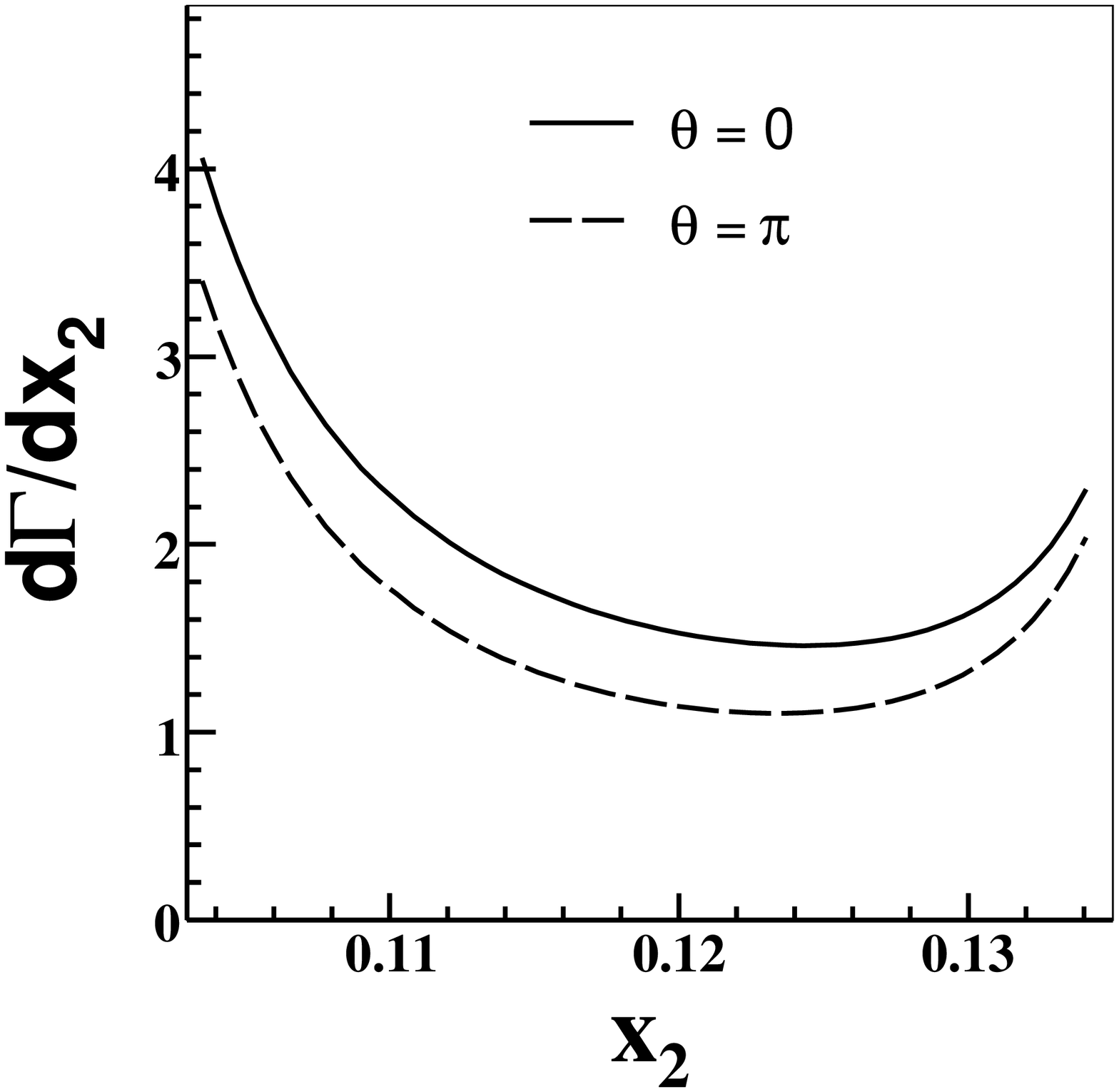}}
\caption{The discrete contributions to the photon
energy spectrum of the $\psi(2S)\to J/\psi+\gamma\gamma$ process in
the cut region: (a) the contribution of the $1P$ states, corresponding to $\Gamma^{1P}_{\rm{Ind}}$ and $\Gamma^{1P}_{\rm{Int}}$ in (\ref{decay}),
(b) the contribution of the $2P$
states and of the interference terms between $1P$ and $2P$ states, corresponding to $\Gamma^{2P}$ and to $\Gamma^{1,2P}_{\rm{Int}}$ in (\ref{decay}),
 (c) the total contribution for a different relative phase angle
$\theta$, corresponding to the $\pm$ sign in (\ref{decay}).} \label{fig:chic_cut}
\end{center}
\end{figure}

The experimental measurement of the $\psi(2S)\to J/\psi+\gamma\gamma$ decay
can be implemented in the off-mass shell region of $\chi_{cJ}$, i.e., the
experimentally sensitive region in Daltiz plot
\begin{equation}
0.15<M_{\gamma\gamma} <0.51\;\mathrm{GeV},\;\;  3.43<M_{J/\psi\gamma} <3.49\;\mathrm{GeV},
\label{eq:cut}
\end{equation}
where $M_{\gamma\gamma}$ is the invariant mass of the two photons and $M_{J/\psi\gamma}$
is the invariant mass of $J/\psi$ and the higher energy photon. These cut can
mostly exclude the contribution from the highly yielded $\chi_{cJ}(1P)$ states.
From the photon spectrum in the cut region indicated in \figurename~\ref{fig:chic_cut},
the different contributions to the decay width
(\ref{decay}) read
\begin{eqnarray}\label{cut1}
\Gamma^{1P}_{\rm{Ind}}=4.68\times 10^{-2}\mathrm{keV}
,\;\;
\Gamma^{1P}_{\mathrm{Int}}=6.5\times10^{-3}\mathrm{keV}\;\nonumber \\
\Gamma^{2P}=1.82\times10^{-4}\mathrm{keV},\;\;
\Gamma^{1,2P}_{\mathrm{Int}}=4.78\times10^{-3}\mathrm{keV}.
\end{eqnarray}
We have also computed the branching ratio, photon spectrum, and the
polarization parameter in the cut region. The result of the branching
ratio is
\begin{eqnarray}\label{cut2}
\mathcal{B}_{\mathrm{dis}}^{\mathrm{cut}}(\psi(2S)\to J/\psi+2\gamma)=\left\{
\begin{array}{ll}
1.92\times10^{-4}, (\rm for\;\; \theta=0)\\
1.60\times10^{-4}, (\rm for\;\; \theta=\pi)
\end{array}
\right.
\end{eqnarray}
The result in Eqs.(\ref{cut1},\ref{cut2}) shows that, contrary to the
case of the total decay width (\ref{total}), in the cut region the effect
of interference among $1P$ states as well as the contribution from the
$2P$ states can not be ignored. It is more than $10\%$ of the
sum of the separated contributions of the $\chi_{cJ}(1P)$ states. The
$J/\psi$ produced in the cut region tend to be unpolarized and the
polarization parameter in the cut region becomes $\alpha^{\mathrm{cut}}=-0.122$
and $-0.107$ for $\theta=0$ and $\theta=\pi$, respectively. If we only
include the three individual contributions of $\chi_{cJ}(1P)$, the value
of $\alpha^{\mathrm{cut}}$ turns to be $-0.078$. Finally, the different
contributions to the photon spectrum in the cut region are shown \figurename~\ref{fig:chic_cut}.

As mentioned before, the coupling constant $\delta^{nP,mS}$ is related
to the spatial matrix element $\langle nP|r|mS \rangle=\int^{\infty}_{0}R_{nP}(r)R_{mS}(r)r^3dr$
in potential models.
At least three different potential models, including the Cornell potential
\cite{Eichten:1974af} and the screened potential \cite{Li:2009zu}, give the
phase angle $\theta$ to be $\pi$ \cite{baiqing}.
Hence, the $\theta=\pi$ option in our calculation appears to be favored.

\section{Discussion and Summary}

We have estimated the discrete contribution to the $\psi (2S) \to
J/\psi + \gamma \gamma$ due to electric dipole transitions in the
whole phase space and, in particular, in the cut region used at BESIII.
Higher multipole electric transitions and the magnetic ones are
suppressed by at least $v_Q^2$ in the amplitude. This is also the
case for contributions arising from two-photon vertices. The largest
uncertainty in our calculation comes from the fact that we neglect
the contribution of $nP$ states for $n\ge 3$. This contribution is
not parametrically suppressed by powers of $v_{Q}$, although we
expect it to be small for the following two reasons: (i) the
propagator of the $nP$ state is increasingly off-shell when $n$
increases, and (ii) the coupling constants $\delta_J^{nP,1S}$ and
$\delta_J^{nP,2S}$ are proportional to the overlap of the radial
wave functions of the corresponding states, which decreases with
$n$. Apart from this truncation, the estimate is reliable at leading
order in $v_{Q}^2$. In fact, the modifications we have made in
(\ref{amplitude}), which introduce terms of higher order in $v_{Q}$,
take into account relativistic effects in the kinematics, and hence
help in providing a better estimate.

From the point of view of potential models, the leading
corrections to the $E1$ transition processes, $\psi(2S)\to
\chi_{cJ}+\gamma$ and $\chi_{cJ}\to J/\psi+\gamma$, mainly arise
from the three following sources: (i) relativistic modification of
the nonrelativistic wave functions, (ii) finite size effects,
and (iii) contribution of high $v_{Q}^2$ order electromagnetic
operators \cite{Eichten:2007qx}. In this work, we use the effective
Lagrangian, Eq.(\ref{amplitude}), to describe the $E1$ transition
process. For the $1P$ case, we determine the values of the
corresponding coupling constants from the experimental data. Thus,
the effects of (i), which are dominant, are taken into account in
our results, as far as the vertices involving the $1P$ states is
concerned. Therefore, our results for the two-photon decay width in
the off-shell region are more accurate than a potential model calculation
for the one-photon $E1$ transition process.
Both the corrections due to
(ii) and to (iii) could be taken into account by including higher
dimensional operators in the vertices (\ref{vertex}). As mentioned
before these operators are suppressed by at least order $v_{Q}^2$.

Let us next discuss how our results compare with the usual inputs in
the Monte-Carlo (MC) codes that are used to analyze the experimental
data. In the experimental treatment of the four contributions in
Eq.(\ref{decay}) in the peaking region, usually only the first one,
$\Gamma^{1P}_{\rm{Ind}}$, is taken into account, and is often modeled
using the nonrelativistic Breit-Wigner line shape of $\chi_{cJ}$ and
$J/\psi$. The other three components are negligible and generally
omitted. But in the off-shell region, as argued above, the other
three components will be sizable and have to be considered in
the data treatment. Furthermore, in the off-shell cut region
(\ref{eq:cut}), even this naive nonrelativistic Breit-Wigner
line shape description of the individual $\chi_{cJ}$ states
contribution needs improvement. The reason is that a (single)
nonrelativistic Breit-Wigner is only a good approximation to the
line shape of $\rm{d}\Gamma^{1P}/\rm{d}x_2$ in the resonance peaking
region, as shown in \figurename~\ref{fig:res_a}. In the cut region
equation (\ref{eq:cut}), which lies between the $\chi_{c0}$ and $\chi_{c1}$
resonance peaks, there is no guarantee that the nonrelativistic
Breit-Wigners will provide a correct description of data. Let us
next point out the main ingredients it misses. In the $E1$
transition process, the decay rate is proportional to the factor
$k_{\gamma}^3$ as shown in Eq.(\ref{E1}). We may then improve on the
nonrelativistic Breit-Wigner approximation by just introducing the
correct photon energy dependence in each vertex i.e. the full
$k_{\gamma 1}^3k_{\gamma 2}^3$ scale factor from the two $E1$
transitions. Note that if one only includes the $k_{\gamma}^3$
correction due to the first $E1$ transition, $\Gamma^{1P}$ in the
cut region will be overestimated, because the energies of the two
photons are negatively correlated.

\begin{figure}
\begin{center}
\includegraphics[width=0.48\textwidth]{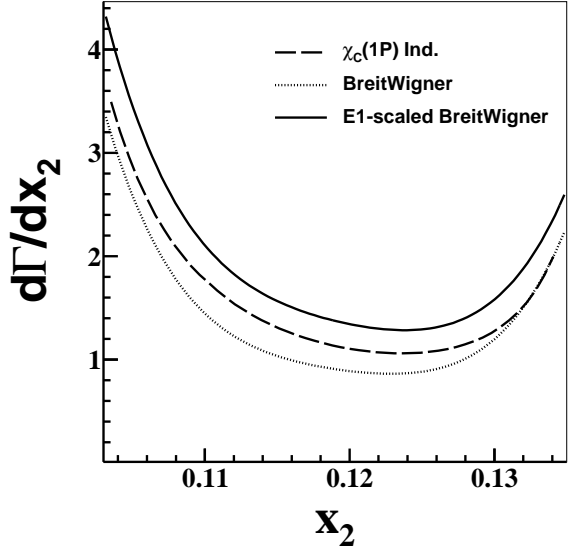}
\caption{The MC simulation of the cascade decay of $\psi(2S)\to
(J\psi\gamma_{1})_{\chi_{cJ}}\gamma_{2}$ in the cut region, where
the branching fractions are from PDG~\cite{Nakamura:2010px}. The
dotted line denotes the naive nonrelativistic Breit-Wigner
simulation, the solid line is the simulation including the
$k_{\gamma_{1}}^3k_{\gamma_{2}}^3$ factor, and the dashed line is the
contribution of the three individual $\chi_{cJ}(1P)$ states, corresponding to $\Gamma^{1P}_{\rm{Ind}}$ in (\ref{decay}), calculated in this paper.
} \label{fig:MC_sim}
\end{center}
\end{figure}

In \figurename~\ref{fig:MC_sim}, the MC simulation results of the
line shapes in the cut region are shown, which are implemented in
two ways. One is done by including the $k_{\gamma 1}^3k_{\gamma
2}^3$ correction from a double $E1$ transition, and the other one is
not. The difference between these two MC simulation results can be
understood as follows: the $E1$ transition enhances the right tail
of photon energy peak and depresses the left tail of the
nonrelativistic Breit-Wigner. As a whole, the effect of the
correlated emitted photons in the double $E1$ transition increases
the $\chi_{cJ}$ contribution in the cut region. For comparison, we
also plot the effective Lagrangian result of the sum of the three
individual $\chi_{cJ}$ states contribution in
\figurename~\ref{fig:MC_sim}. It is clear that neither of the
simulation results agree with that of the effective Lagrangian
calculation, although they are qualitatively similar. From the
amplitude in Eq.(\ref{amplitude}), it can be verified that the
$k_{\gamma 1}^3k_{\gamma 2}^3$ factor and the nonrelativistic
Breit-Wigner is only the leading-order nonrelativistic
approximation of the effective Lagrangian calculation in the
off-shell region. Hence, even for a
delicate description of the individual contribution in the off-shell
region, including the double $E1$ transition correction  in the
naive nonrelativistic Breit-Wigner only may not be enough.


In summary, we have estimated the discrete contribution to the $\psi
(2S) \to J/\psi+\gamma\gamma$ due to electric dipole transitions in
the whole phase space, and in particular in the cut region in the
experimental measurement.
We find that for the full decay width the interference contribution and the
contributions of higher excited states can be safely neglected. However,
in the regions of the phase space off the resonance peaks,
their contributions are considerable and
important for a delicate experimental measurement. As argued in the
Introduction, a large deviation of our results from an
experimental observation, would indicate that the effects of the
$D\bar D$ threshold are significant.


\section*{Acknowledgments}
We would like to thank Bai-qing Li for correspondence about the
relative phase between $1P$ and $2P$ contribution. He and Lu also
thank De-Shan Yang and Qiang Zhao for helpful discussions. He and J.S.
are supported by the CSD2007-00042 Consolider-Ingenio 2010 program
(He under Contract No.CPAN08-PD14) , and by the FPA2007-66665-C02-01/
and FPA2010-16963 projects (Spain). J.S. has also been supported by
the RTN Flavianet MRTN-CT-2006-035482 (EU), the ECRI HadronPhysics2
(Grant Agreement No. 227431) (EU), Grant No. FPA2007-60275/MEC
(Spain) and the CUR Grant No. 2009SGR502 (Catalonia). This work is also
supported in part by National Natural Science Foundation of China
(NSFC) 10905091, 100 Talents Program of CAS, Ministry of Science and
Technology of China (MOST), Foundation B of President of GUCAS, SRF
for ROCS of SEM, and China Postdoctoral Science Foundation.

\end{document}